\documentclass[aps,pre,amsmath,amsfonts,floatfix]{revtex4}

\usepackage{graphicx}

\font\msytw=msbm9 scaled\magstep1

 \let\b=\beta    \let\d=\delta

 \let\t=\tau   \let\f=\varphi 
   
 \let\D=\Delta   
    \let\Si=\Sigma     
  \let\r=\rho 
\let\io=\infty
\def\ie{{i.e. }}\def\eg{{e.g. }}

\def\PPP{{\cal P}}

\def\erf{\text{erf}}

\def\to{\rightarrow}

\def\RRR{\hbox{\msytw R}}

\newcommand{\beq}{\begin{equation}}
\newcommand{\eeq}{\end{equation}}

\begin{document}

\title{Some recent theoretical results on amorphous packings of hard spheres}

\author{Francesco Zamponi\footnote{francesco.zamponi@phys.uniroma1.it}}

\affiliation{Laboratoire de Physique Th\'eorique de l'\'Ecole Normale Sup\'erieure,
24 Rue Lhomond, 75231 Paris Cedex 05, France}

\date{\today}

\begin{abstract}
The aim of this paper is to
review and discuss qualitatively some results on the properties of amorphous
packings of hard spheres that were recently obtained by means of the replica method.
The theory gives predictions for the equation of state of the glass, the
complexity of the metastable states, the scaling of the pressure 
close to jamming, the coordination of the packing and the pair correlation 
function in any space dimension $d$. 
The predictions compare very well with numerical simulations in $d=2,3$.
The asymptotic predictions for $d\to\io$ are within the rigorous bounds.
The theory can be extended to binary mixtures and to hard core potentials with
an attractive tail or square well.
\end{abstract}


\maketitle

The study of amorphous packings of hard spheres is relevant for a large class
of problems, including liquids, glasses, colloidal dispersions, granular
matter, powders, porous media, etc. and a large amount of numerical and
experimental data is available in the literature
\cite{Be83,SK69,Fi70,Be72,Ma74,Po79,Al98,SEGHL02,To95,RT96,Sp98,Torquato,HLLN02,DTS05,Luca05,TTD00,XBH05}.
Moreover, the problem of sphere packing is related to many mathematical problems
and arises in the context 
of signal coding and of error correcting codes, therefore it has been investigated
in detail by the information theory community \cite{ConwaySloane,Rogers}.
Nevertheless, a satisfactory characterization of the amorphous states of a system of
identical hard spheres is not yet available and in particular 
the question whether a glass transition exists 
in finite dimension is still open.
From a rigorous point of view, for space dimension $d>3$ 
only some not very restrictive bounds on the maximal density have been obtained,
and in particular it is still unclear whether
the densest packings for $d\to \io$ are amorphous or crystalline.

Recently, a quantitative description of the glass transition in structural glasses has been obtained
by means of the replica trick \cite{MP99,MP99b,MP00,CMPV99,CFP98,PZ05}. This method was
successfully applied to Lennard-Jones systems \cite{MP99,MP99b,MP00,CMPV99} and, more recently, to
hard spheres in $\RRR^d$ \cite{CFP98,PZ05,PZ06a},
on the Bethe lattice \cite{PTCC03}, and on the hypercube
\cite{PZ06b}.
In this approximation the glass transition turns out to be similar to the 1-step replica symmetry
breaking (1RSB) transition that happens in a class of mean-field spin glass models and indeed the
replica strategy described above was inspired by the exact solution of these models
\cite{MPV87,KTW87b,GM84,Mo95}.  

In this paper I review these results, for the case of hard spheres in $\RRR^d$,
without discussing the replica method, for which the reader can refer to the original
papers \cite{MP99,MP99b,MP00,PZ05,PZ06a}:
here I will discuss in details the limits of the theory and how it compares with numerical results.
As a matter of fact, despite the strong idealizations involved in the theory, the agreement with
numerical data is surprisingly good. 

\section{What is an ``amorphous packing''?}

In this section we will try to define ``amorphous packings'' of hard sphere systems and
in particular the ``random close packing'' density, \ie the maximum density of amorphous packings.
We will see that there are difficulties in the theoretical definition of this concept that allow
only to define it {\it approximately}.
The usual way to estimate this quantity in experiments or numerical simulation is to compress
the system according to some protocol: 
\eg in numerical simulations the particle diameter is
slowly increased during a molecular dynamics run, or in experiments particle are thrown 
randomly in a box and the box is subsequently shaked
\cite{Be83,SK69,Fi70,Be72,Ma74,Po79,Al98,SEGHL02,To95,RT96,Sp98,Torquato,DTS05,TTD00,XBH05}. 
The density of the system increases with time and
usually approaches a value of $\f_{RCP} \sim 0.64$.
However the precise value of the latter quantity is found to depend in a non trivial way on the 
details of the experimental protocol.
Then, to find a theoretical estimate of $\f_{RCP}$ we will have to refer to some 
idealization. For concreteness we will discuss the problem in $\RRR^3$ but the discussion applies
to any space dimension.

\subsection{The fate of the liquid above freezing: a naive definition of random close packing}

\begin{figure}
\includegraphics[width=8.5cm]{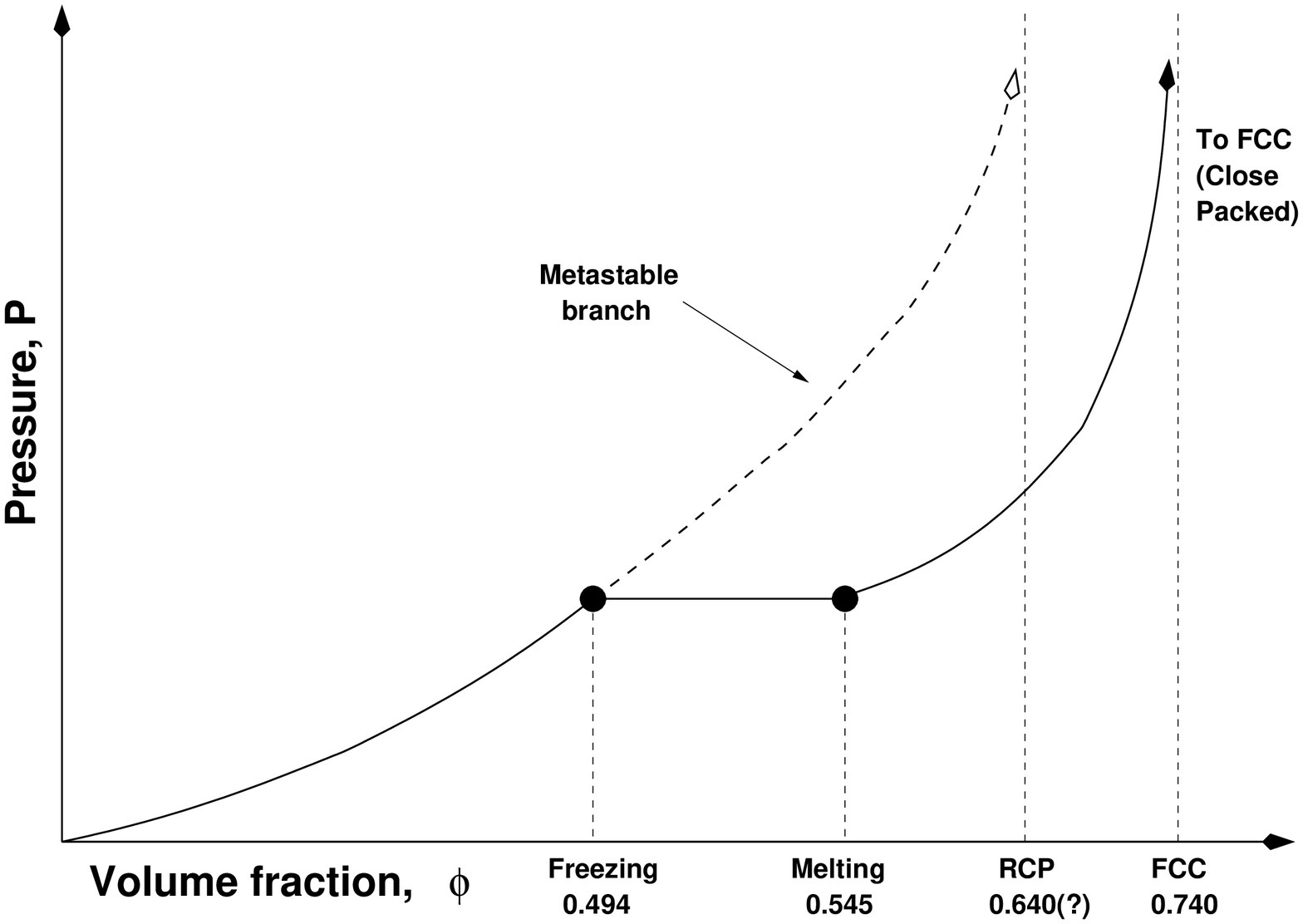}
\includegraphics[width=8.5cm]{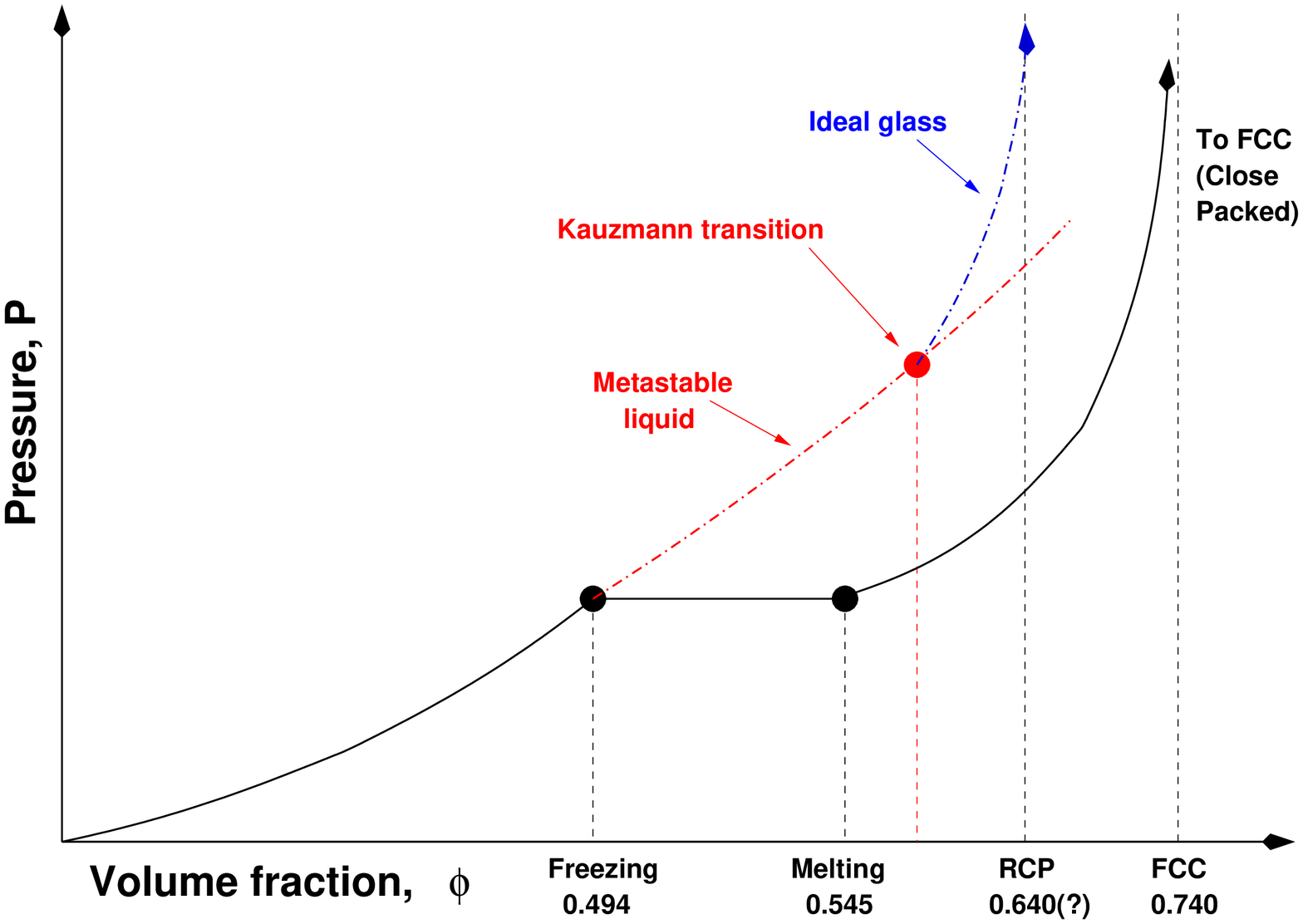}
\caption{Schematic phase diagram of hard spheres in $\RRR^3$. (Left)
  Continuation of the liquid equation of state in the metastable
  region. (Right) Expected behavior in presence of a thermodynamic glass transition.}
\label{fig:dialiquido}
\end{figure}

The equilibrium phase diagram of hard spheres in $\RRR^3$ is sketched in Fig.~\ref{fig:dialiquido}, where
the density is reported as a function of the {\it packing fraction} $\f=\r V_d(D/2)$, $D$ being
the diameter of a sphere and $V_d(r)$ the volume of a $d$-dimensional sphere of radius $r$,
so that $\f$ is the fraction of volume covered by the spheres. At low density
the system is in a liquid phase (as defined \eg by the low density virial expansion); the maximum
possible density is realized by the FCC lattice. A first order phase 
transition between the liquid phase and the FCC crystal phase is found by numerical simulations.

The first naive idea to define amorphous states of hard spheres at high density is to assume
that the liquid phase can be continued above the freezing density $\f_f$ and to look at its properties
at large density. For instance one can choose a functional form that represents well 
the equation of state of the liquid below $\f_f$ (\eg the Carnahan-Starling or Percus-Yevick equation of 
state~\cite{Hansen}) and assume that it describes the liquid phase also above $\f_f$.
On increasing the density the pressure of the liquid increases as the
average distance between particles decreases: one may expect that it 
diverges at a point where the particles get in contact with their neighbors,
and the system
cannot be further compressed. Then one can call this point {\it random close packing density} (RCP),
see left panel of Fig.~\ref{fig:dialiquido}.

The first objection to this proposal is that an intrinsic stability limit of the liquid 
(a spinodal point) might exist at a density above $\f_f$ due to thermodynamic or kinetic
reasons. A thermodynamic spinodal is likely not to be there and any reasonable continuation
of the liquid equation of state do not predict such an instability (manifested \eg by an infinite
compressibility). A kinetic spinodal, related to the existence of the crystal
\cite{CAL05}, could instead exist, at least in monodisperse systems. This
would imply the impossibility of reaching amorphous jammed states if the
compression rate is not very high.
We will assume in the following that the metastable liquid can be compressed
as slow as wished avoiding crystallization. This is not a very good
assumption for monodisperse systems but seems to be more close to reality for a
suitably chosen binary mixtures. 
This point requires a better investigation, \eg following the analysis of \cite{CAL05},
and we will not discuss it further here.

A second objection is that in presence of a first order phase transition the
continuation of one phase into its metastable region is not well defined due to the appearance
of essential singularities at $\f_f$. Many possible continuations of the
low density equation of state above $\f_f$ are possible: this means that the properties
of the ``liquid'' above $\f_f$ will depend on the history of the sample, much as it happens for
the hysteresis of a ferromagnetic system. This means that, even neglecting
crystallization, the definition of RCP above is not precise from a theoretical point of view.
Nevertheless, the ambiguity is expected to be exponentially small in the distance from $\f_f$,
and as the distance between $\f_f$ and the maximum density $\f_{FCC}$ is not so large, one might
expect to obtain a meaningful result in this way. And indeed the difference between different possible
continuation of the liquid equation of state (\eg Carnahan-Starling, Percus-Yevick, Hypernetted Chain)
is of the order of $10\%$ above $\f_f$. 

The main problem with this definition is that all these equations predict a divergence of the pressure
at unphysical large values of $\f$: \eg the CS equation predicts $\f_{RCP} = 1$ which is clearly
wrong since it is larger than the FCC value and moreover implies that the available volume is 
completely covered by the spheres. Thus this idea is definitely too naive and we have to resort
to something more refined.

\subsection{The ideal glass transition: a (still naive) definition of RCP}

A way out of this contradiction is to assume the existence of a thermodynamic glass transition in
the {\it metastable} liquid branch of the phase diagram.
Such a transition can be expected for various reasons
but we will not try to justify it: we will simply assume that it is there and investigate the
consequences of such an assumption.

So let us assume again that the liquid phase can be continued above $\f_f$ and neglect the
(small) ambiguity in its definition due to its intrinsic metastability with respect 
to crystallization. We assume that at a density $\f_K$ a {\it thermodynamic glass transition} 
(sometimes called {\it ideal glass transition}) happens.
The transition is signaled by a jump in the compressibility of the system. A
simple qualitative argument
to explain this is the following: in the dense liquid phase particles vibrate on a fast time
scale in the cages made by their neighbors, while on a much larger time large scale
cooperative relaxation processes happen ({\it structural relaxation}). 
If we change the density by $\D \f$ the pressure
will instantaneously increase by a $\D P_0$; very rapidly the average size of the cages will
decrease a little due to the increase in density and the pressure will relax to a value
$\D P_f < \D P_0$. Then, on the time scale of structural relaxation, the structure will change
to follow the change in density and the pressure will relax further to a value $\D P_\io < \D P_f$.
At the glass transition the latter relaxation is frozen as the corresponding time scale becomes
infinite. Thus, in the glass phase the increase in pressure following a change in density will
be larger than in the liquid phase, leading to a smaller compressibility $K = \f^{-1} (\D \f/\D P)$.

The schematic phase diagram that we expect in presence of a glass transition is in 
right panel of Fig.~\ref{fig:dialiquido}.
It is evident that the existence of a glass transition can cure the paradoxical behavior
of the pressure of the liquid, that seems to diverge at a density bigger than the FCC density.
Then we can define the RCP density as the point where the pressure of the {\it glass} diverges.

Still this definition of RCP in unsatisfactory for two reasons. 
The first is that the ambiguity in the definition of the liquid equation of
state due to its metastability affects also the glass: so the glass equation
of state will be {\it theoretically} not well defined, with an ambiguity of
the order of $10\%$, depending on the equation of state one chooses to describe
the liquid.
Moreover, the existence of the crystal means
that one can construct configurations of the system representing mixtures of the crystal and the glass
states; such configurations will have an arbitrary degree of local order and will span the whole
range of densities between the density of the glass and that of the crystal: explicit examples have
been constructed in \cite{TTD00}. It is very difficult to define an ``order metric'' to quantify
the order in a given configuration of a large system. Thus {\it on a practical ground} it is very
difficult to distinguish a ``pure glass state'' from a mixture of glass and crystal. To resume,
the existence of the crystal poses both theoretical and practical difficulties in defining RCP
as the point where the pressure of the equilibrium glass diverges.

A second difficulty is the following. The glass transition, in the standard picture coming from
the analysis of mean field models, is due to the appearance of many metastable states: in the case
of hard spheres this means that we expect that, in addition to the stable glass states corresponding
to the equilibrium glass, we will have many {\it metastable packings}.
These packings can be described as follows: in the equilibrium liquid at a given (high) density
$\f < \f_K$ the particles will vibrate around local stable structures, that are visited subsequently
on the scale of the structural relaxation. If we now ``artificially'' froze the structural relaxation
and compress the system, the pressure will increase faster than if the structure is allowed to relax
for the same reason as above: the system will be forced to reduce the size of the cages to respond
to a change in density. Then the pressure will diverge at the point where the particles get in contact
with their neighbors and the average size of the cages is zero. In this way we can produce {\it jammed}
configurations in a whole range of densities $\f < \f_{RCP}$ \cite{TTD00,Torquato,DTS05,XBH05}.
Experimentally or numerically this can be done by compressing fast enough in order to disallow the
system to relax the structure during compression. And it is clear that, as the time scale for
structural relaxation is expected to diverge on approaching $\f_K$, at some point it will
fall beyond any experimentally accessible time scale and necessarily the system will be frozen
into a metastable state. Thus the {\it ideal} glass states are likely to be unobservable in
practice: one will observe instead a {\it nonequilibrium} glass transition to a state that, again,
will depend on the experimental protocol.

To summarize, the difficulties in the definition of the random close packing
are mainly due to {\it metastability}, either of metastable glass states with
respect to the ideal glass states, or of the whole glass states with respect to the crystal.
Thus, in the following we will {\it assume} the idealized phase diagram given in
Fig.~\ref{fig:dialiquido}, neglect the existence of the crystal and try to compute properties 
of the pure ideal glass state, and possibly also of metastable glass states. 
We will see that even in this ``zeroth order'' approximation we can obtain a very good 
agreement with numerical simulations. Moreover, we will consider the limit
$d\to\io$ in which the ambiguities related to metastability are expected to be
much smaller. We believe that it is possible that in this limit the theory is exact.

The method we used has been described in \cite{PZ05,PZ06a}. In an extended version of this
paper we will present a simpler derivation, explain how crystallization is
removed in the theory and qualitatively discuss the effect of 
metastability.

\subsection{On the protocol dependence of the random close packing density}

Generally speaking, we can use a given algorithm to produce jammed configurations of
hard spheres: the case we discussed above corresponds, for
instance, to a molecular dynamics simulation during which the system is compressed 
at a given rate. The algorithm will stop when the system is jammed: the final
density will be a random variable depending on the initial data and possibly
on some randomness built in the algorithm itself. We can define the
probability $\PPP(\f_J)$ of reaching a final density $\f_J$: it will be the
product of the {\it density of states} $\r(\f_J)$, \ie the number of jammed
configurations with density $\f_J$, times the probability $\beta(\f_J)$
that the algorithm finds a particular configuration with density
$\f_J$. The former quantity depends only on geometrical
properties of the configuration space of hard spheres, while the latter encodes the properties
of the algorithm~\cite{XBH05}.

We can compute the {\it complexity}
$\Si(\f_J) = \lim_{N\to\io} N^{-1} \log \rho(\f_J)$ for the ``amorphous
states'' defined above. This quantity is reported
in Fig.~\ref{fig:diad3}. It vanishes linearly at $\f_{RCP}$ giving an
exponential distribution of the jammed states, 
$\rho(\f_J) \sim \exp[N(\f_J-\f_{RCP})]$. However the quantity $\beta(\f_J)$
is also expected to scale as $\b(\f_J) \sim \exp[N b(\f_J)]$; the
resulting $\PPP(\f_J)$ will be strongly peaked around a given value $\f_J^*$
-- the maximum of $\Si(\f_J)+b(\f_J)$ -- \cite{Kuunp},
which however might depend on the particular algorithm through the function
$\b(\f_J)$, giving rise to the protocol dependence observed in experiments,
see \cite{XBH05} for a detailed discussion of this important issue. However,
the value $\f_{RCP}$ defined as the point where $\Si(\f_J)$ vanishes is a
property of the system that does not depend on the algorithm (at least if one
accepts the idealizations discussed in the previous subsection, \ie if one
neglects the existence of the crystal).

\section{Results}

\subsection{Equation of state of the glass}

The only input of the theory is the entropy of the
liquid, $S(\f)$. As discussed above, this quantity is not well defined above
the freezing density and we have to choose an expression which describes well
the liquid below $\f_f$ and extrapolate above $\f_f$.

\vskip5pt

\noindent {\bf d=1} -
In dimension $d=1$ we do not expect any glass transition. 
The entropy can be computed exactly and is given by 
$S(\rho)= 1 - \log \rho + \log (1-\rho D)$:
it diverges at the close packing $\rho D=1$. Indeed we find no glass transition.

\vskip5pt

\noindent {\bf d=2} -
In $d=2$ we can use either the Henderson expression for $S(\f)$ \cite{He75},
or the improved expression of Luding \cite{Lu01},
with very small quantitative differences.
Again we find no glass transition, consistently with the numerical simulations
of \cite{SK00}, but we find a glassy solution which never becomes stable. 
The properties of the amorphous states
seem to be similar to the one observed in the numerical simulations of \cite{BW06} (see below).
However in $d=2$ the hard sphere liquid has very peculiar properties so this case
deserves further investigation.

\begin{figure}
\includegraphics[width=8cm]{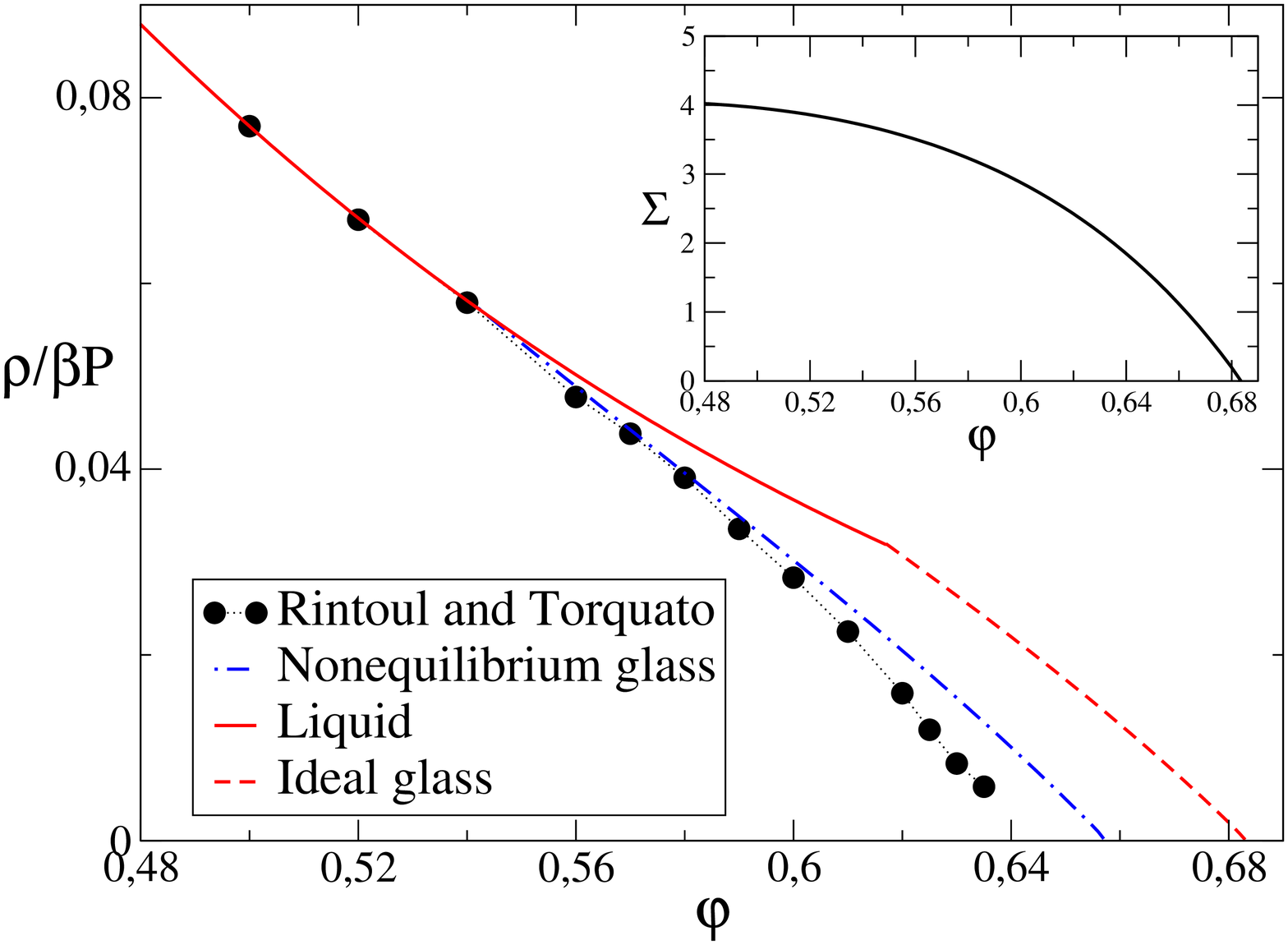}
\includegraphics[width=8cm]{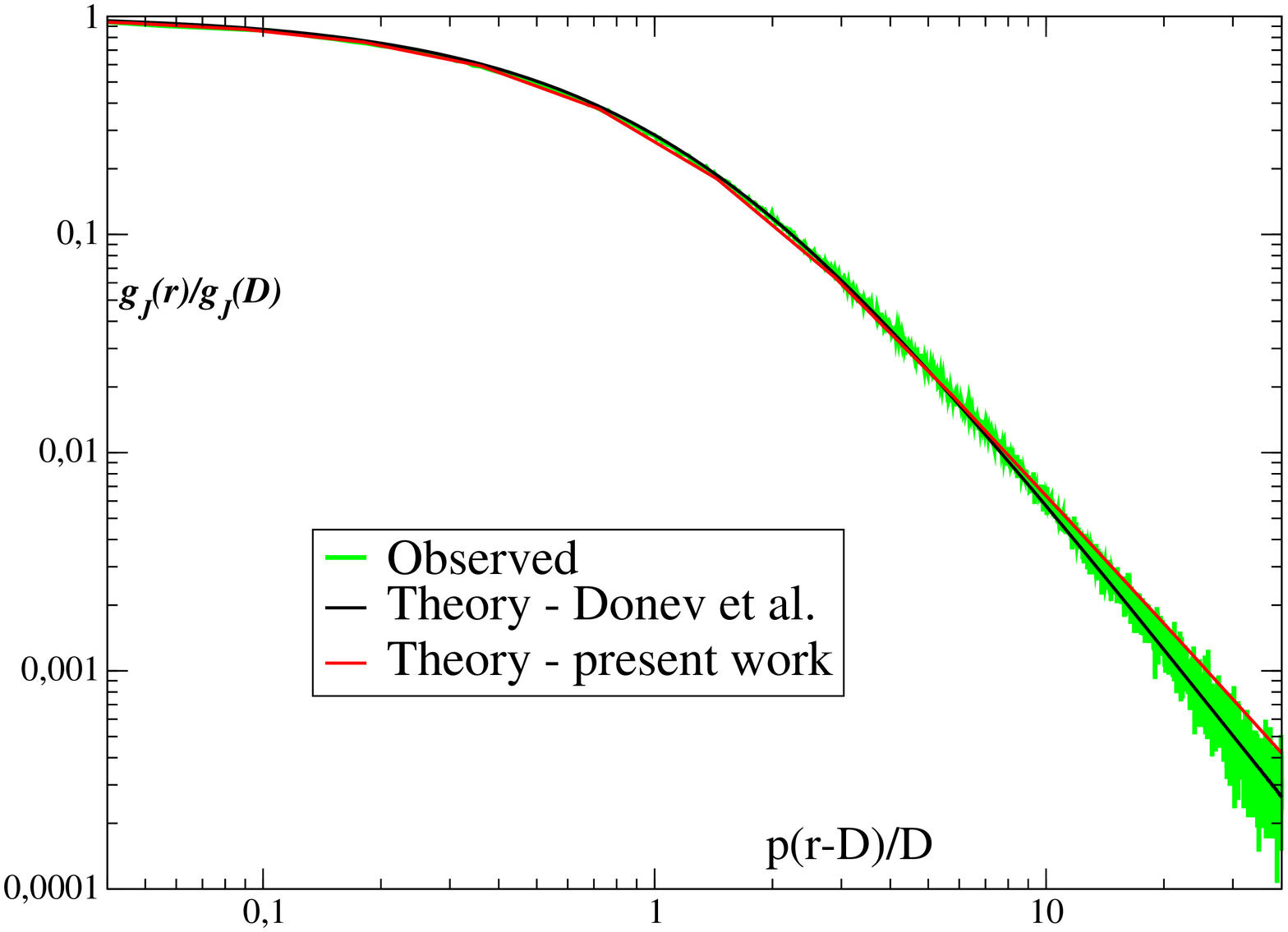}
\caption{(Left) Equation of state in $d=3$ using the Carnahan-Staling equation of
  state to describe the liquid. The inverse of the reduced pressure $\b P/\r$ is reported 
as a function of $\f$.
The ideal glass transition density is $\f_K =
  0.62$ and the random close packing density is $\f_{RCP} = 0.68$. The metastable
  state reported in the figure correspond to a jamming
density $\f_J =0.658$. Numerical data from \cite{RT96} are reported for comparison.
(Inset) Complexity as a function of the jamming density for $d=3$.
(Right) Scaled correlation function $g_J(r)/g_J(D)$ close to $r=D$. The theory
is compared with numerical data from \cite{DTS05}.
}
\label{fig:diad3}
\end{figure}

\vskip5pt

\noindent {\bf d=3} - 
In $d=3$ we used the Carnahan-Starling expression \cite{Hansen} for the entropy, 
which reproduces very well the numerical data for the equation of state.
We predict a glass transition density $\rho_K \sim 0.62$
in good agreement with numerical results. The pressure of the equilibrium glass
diverges at $\f_{RCP} \sim 0.68$. We can also compute the equation of state of
the metastable (or nonequilibrium) glasses. The latter are labeled by their jamming density
$\f_J$ which is the density at which the pressure of the metastable glass diverges.
We find a whole distribution of metastable glasses with jamming densities $\f_J < \f_{RCP}$.
A comparison with numerical data is in Fig.~\ref{fig:diad3}: the comparison
suggests that the value $\f_{J} \sim 0.64$ observed in numerical simulation
correspond to the jamming density of a {\it metastable} state in which the
system is frozen due to the finite accessible time scale. The {\it ideal}
value of $\f_{RCP} \sim 0.68$ would be reached only by an infinitely slow
compression (if crystallization is avoided).

\vskip5pt

\noindent {\bf d$\to\io$} -
The entropy of the hard sphere liquid for
$d\to \io$ was computed in \cite{FP99,PS00} in two different ways,
and in both cases it was found that $S(\f)$ is given by the ideal gas term plus
the first virial correction (\ie by the Van der Waals equation),
$S(\f) = 1-\log \r - 2^{d-1}\f $
with exponentially small correction in $d\to\io$. Using this expression we
find that both $\f_K$ and $\f_{RCP}$ scale asymptotically as $d \log d / 2^d$,
with $\f_{RCP}-\f_K \sim d/2^d$.
We can compare this prediction for the maximum density of amorphous packings
with the best available bounds on the density of crystalline
packings. 
Unfortunately, the best lower bound for periodic packings is the Ball bound $\f > d/2^d$,
while the best upper bound is the Blichfeldt's one, $\f < 2^{-d/2}$
\cite{ConwaySloane,Rogers}. Our result for $\f_{RCP}$ lies between these bounds so
we cannot give an answer to the question whether the densest packings of hard spheres
in large $d$ are amorphous or crystalline. Hopefully better bounds on the density
of crystalline packings will address this question in the future. Recent
related work appeared also in \cite{Tordlarge}.

\subsection{Scaling close to jamming}

We can derive asymptotic relations for the behavior of the metastable states
at $\f \to \f_J$ (the equilibrium glass corresponds to the particular case $\f_J=\f_{RCP}$). 

The complexity $\Si(\f_J)$,
that represents the logarithm (per particle) of the number of jammed structures 
with density $\f_J$, is given by
\beq
\Si(\f_J) = S(\f_J) -d \log \left[\frac{\sqrt{8}}{2^d \f_J Y(\f_J)}\right] + 
\frac{d}2 \ ,
\eeq
where $Y(\f_J)$ is the value of the pair correlation at contact. It is reported
in Fig.~\ref{fig:diad3} for $d=3$. Note that unfortunately we cannot investigate the
stability of the {\it metastable} packings so we do not know the minimal value of $\f_J$.

The entropy of a metastable glass is found to diverge as
\beq
s(\f,\f_J) \sim d \log(\f_J-\f)
\eeq
close to $\f_J$, so that the (reduced) pressure diverges as
\beq
p(\f,\f_J) = -\f \frac{d s(\f,\f_J)}{d\f} \sim \frac{d \ \f_J}{\f_J-\f} \ .
\eeq
It is important to remark that the corrections to the leading order seem to be quite large, as observed in
numerical simulations.

\begin{figure}[t]
\includegraphics[width=8cm]{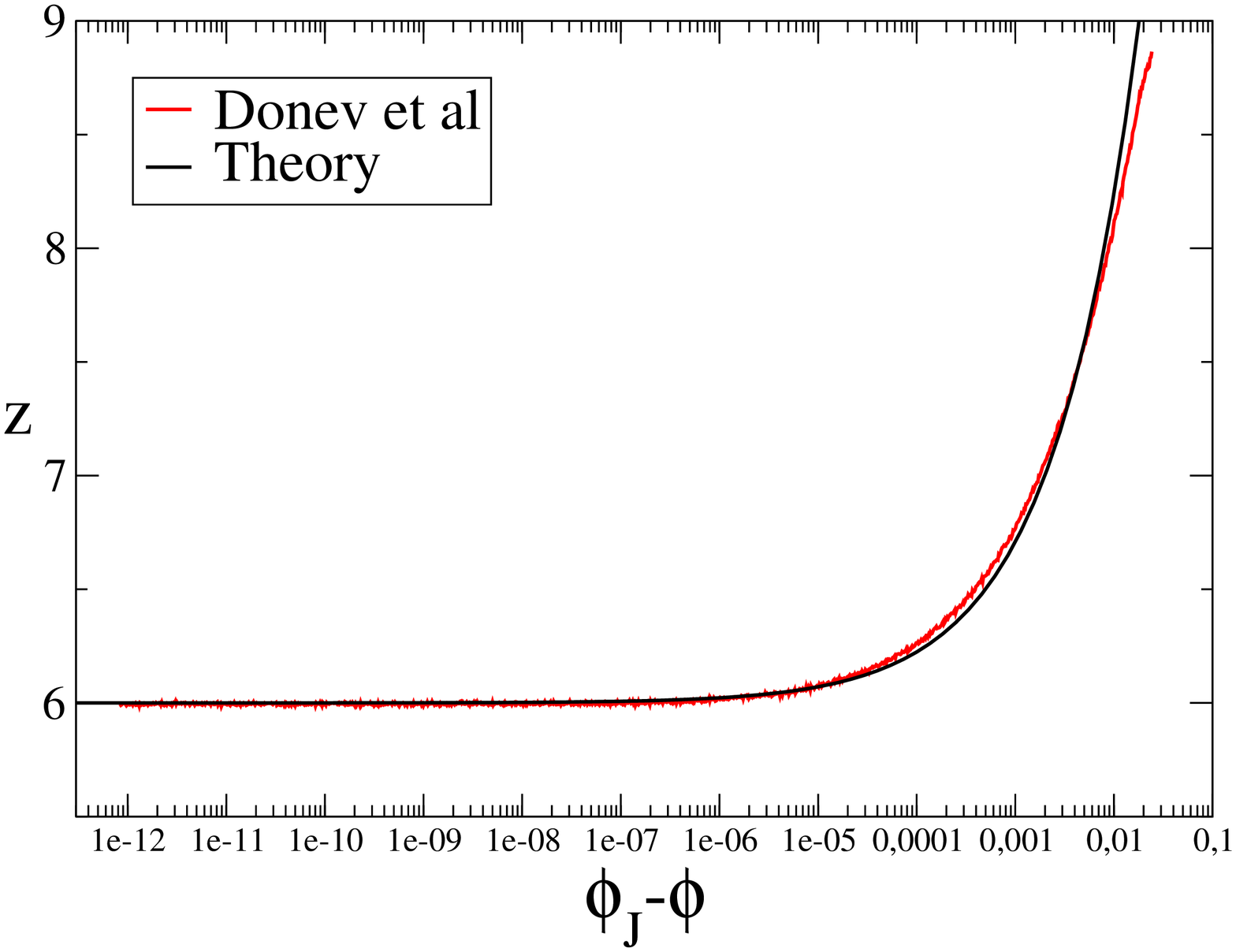}
\includegraphics[width=8cm]{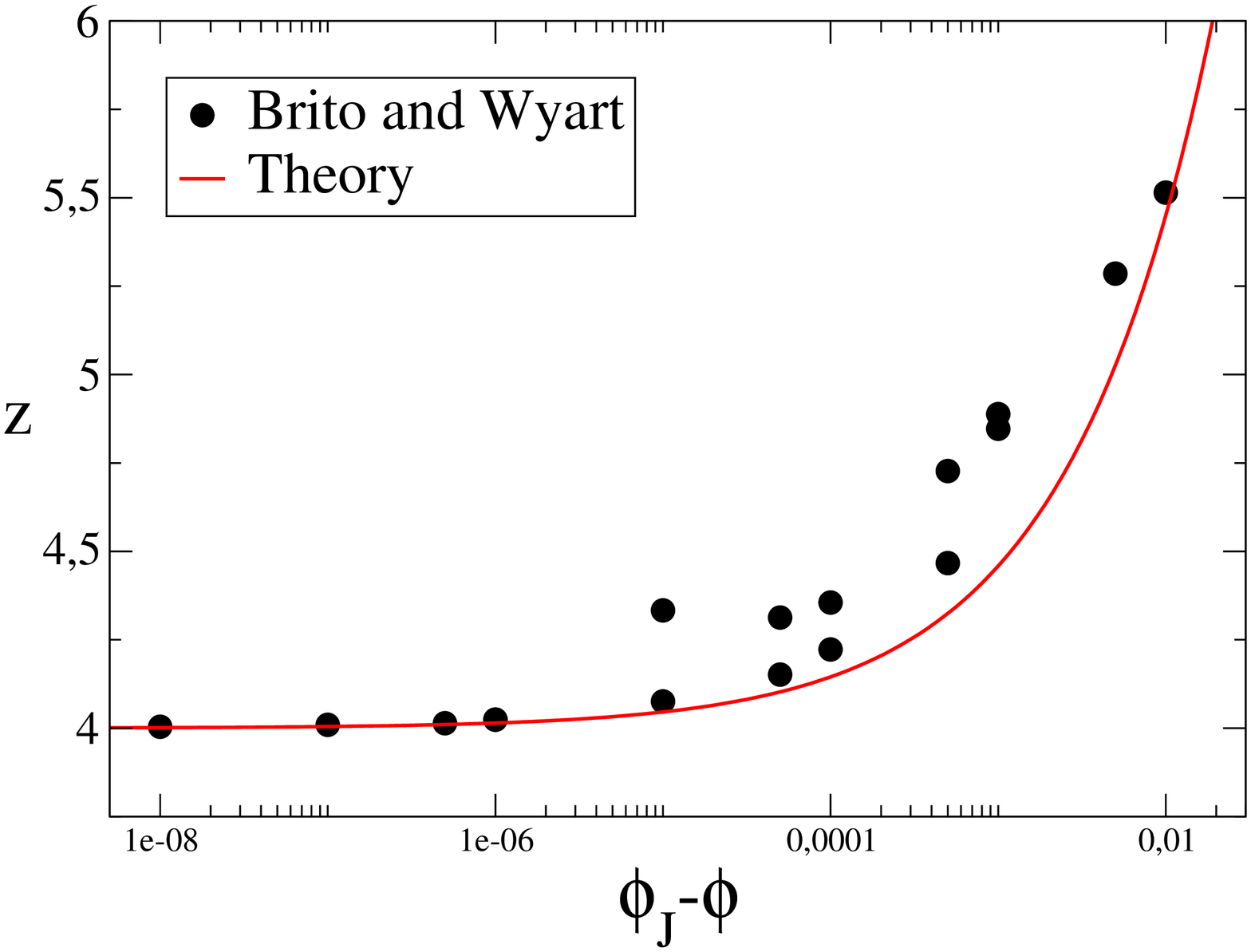}
\caption{
Number of contacts. (Left) Data from \cite{DTS05} in $d=3$ and $\f_J \sim
0.64$. The theory predicts $z = 6 + 20 \sqrt{\f_J-\f}$.
(Right) Data from \cite{BW06} in $d=2$ and $\f_J \sim 0.83$.
The theory gives $z=4 + 14.5 \sqrt{\f_J-\f}$. 
}
\label{fig3}
\end{figure}

\subsection{Correlation function}

The most interesting results on the scaling close to $\f_J$ are obtained by investigating
the pair correlation function $g_J(r)$ of the spheres in the packing. 
This quantity can be (partially) computed within our theory: the details are in \cite{PZ05}.
We find that $g_J(r) = g(r) [1+\d g_J(r)]$, where $g(r)$ is the pair correlation of the liquid
and $\d g_J(r)$ is a correction which is large for $r-D \underset{\sim}< \sqrt{\f-\f_J}$.
In this region we have, for $\f \to \f_J$:
\beq\label{deltapeak}
\frac{g_J(r)}{g_J(D)} = \D_0 \left[ \frac{ \sqrt{\pi}}{2} \,\frac{ p(\f,\f_J) (r-D)}{D} \right] \ , \hskip30pt 
\D_0(u) = 2 \int_0^\io dy \, y \, e^{-y^2 - 2 u y} 
= 1 - \sqrt{\pi} u e^{u^2} (1 - \erf(u))
\ .
\eeq
In Fig.~\ref{fig:diad3} we report numerical data from \cite{DTS05} on the delta-peak contribution
for a packing in $d=3$ with $\f_J \sim 0.64$ and $\f_J-\f \sim 10^{-12}$.
We see that our result, given by Eq.~(\ref{deltapeak}) for $\f_J=0.64$,
perfectly agrees with the data. Unfortunately there are some contributions to
$g_J(r)$ that we still are not able to compute: we expect these corrections to
be responsible for the power law divergence of $g_J(r)$ for $r \to D$ and for
the splitting of the second peak as described \eg in \cite{DTS05}.

\subsection{Number of contacts}

For $\f \to \f_J$ there is a
shell of width $\sqrt{\f_J -\f}$ around a given particle where the probability of finding
other particles is very high, as expressed by the correction $\d g_J(r)$.
These particles will become {\it neighbors} of the particle in the origin
at $\f=\f_J$, and their number is finite as 
the integral of $\d g_J(r)$ on this shell is finite.
The integral of $g_J(r)$ on a shell $D \leq r \leq D + O(\sqrt{\f_J-\f})$ gives
then the average number of contacts for $\f \to \f_J$, 
which turns out to be $z = 2 d [1 + A(\f_J) \sqrt{\f_J-\f}]$, where the coefficient
$A(\f_J)$ can be explicitly computed.
For $\f_J - \f$ finite but small this number can be interpreted as the number
of particles that collide with the particle in the origin during a finite but
long time $\t$ \cite{DTS05,BW06}.
The result is compared with numerical data in Fig.~\ref{fig3}. Remarkably it compares
well with simulations also in $d=2$, indicating that the (unstable) glass phase we find in $d=2$
might have some connection with bidimensional 
amorphous packings.

\section{Conclusions}

The replica method seems to be a powerful tool to investigate the properties
of amorphous packings of hard spheres. Despite the idealizations involved in
this approach, the results seem to compare well with numerical simulations in
$d=2,3$. As the method works in any space dimension, it gives predictions in
$d>3$ that we hope will be tested in the future. In the limit $d\to\io$ the
method is likely to become exact and we hope that this work will stimulate new
more rigorous investigations of this (and related) problems.

There are still a number of issues that have to be addressed within the
theory: for instance, a full computation of the pair correlation of the glass
is mandatory, and an investigation of the stability of the jammed
configurations should be performed.

Work is in progress to extend the method to the case of binary mixtures,
following \cite{CMPV99}. In this case crystallization should be less relevant
and more reliable prediction should be obtained, as in the case of
Lennard-Jones systems. We would like also to apply the method to hard
spheres with attractive potentials like Yukawa tails or square wells: for
these potentials Mode-Coupling theory predicts a rich phenomenology
(reentrance of the glass transition line, glass-to-glass transition, ...)
\cite{Zac}. Some of these results have indeed already been reproduced within a
replica method similar to the one presented here \cite{Velenich}.

\acknowledgments

I would like to thank the organizers of the X International Workshop on
Disordered Systems for the opportunity to present these results. 
The paper is based on a joint work with G.~Parisi~\cite{PZ05,PZ06a}, 
and I wish to thank him for
a careful reading of the manuscript. I would also like to thank the authors of
\cite{RT96,DTS05,BW06} for the kind permission to reproduce their numerical data.

This paper has been motivated by many discussions with G.~Biroli, J.-P.~Bouchaud,
A.~Cavagna, B.~Coluzzi, A.~Donev, S.~Franz, A.~Giuliani, C.~O'Hern, J.~L.~Lebowitz,
R.~Monasson, W.~Krauth,
J.~Kurchan, F.~Sciortino, F.~H.~Stillinger, M.~Tarzia, S.~Torquato,
P.~Verrocchio, M.~Wyart:
I wish to thank all of them for their interest in our work and for their suggestions.

I benefited a lot from the stimulating atmosphere of the Department of
Physics of the Princeton University where part of this work was developed, and
from the discussions at the end of my presentations in LPTHE Jussieu, ENS
Paris and Princeton. I wish to thank the organizers for the invitation and 
all the participants for their comments.

This work has been supported by the Research Training Network STIPCO (HPRN-CT-2002-00319).


\begin{thebibliography}{99}

\bibitem{Be83} J.~G.~Berryman, Phys.~Rev.~A {\bf 27}, 1053 (1983).

\bibitem{SK69} G.~D.~Scott and D.~M.~Kilgour, 
Brit.~J.~Appl.~Phys. (J.~Phys.~D) {\bf 2}, 863 (1969).

\bibitem{Fi70} J.~L.~Finney, Proc.~R.~Soc.~London, Ser.~A {\bf 319},
479 (1970).

\bibitem{Be72} C.~H.~Bennett, J.~Appl.~Phys. {\bf 43}, 2727 (1972).

\bibitem{Ma74} A.~J.~Matheson, J.~Phys.~C:~Solid State Phys. {\bf 7}, 2569 (1974).

\bibitem{Po79} M.~J.~Powell, Phys.~Rev.~B {\bf 20}, 4194 (1979).

\bibitem{Al98} S.~Alexander, Phys.~Rep. {\bf 296}, 65 (1998).

\bibitem{SEGHL02} L.~E.~Silbert, D.~E.~Ertas, G.~S.~Grest, T.~C.~Halsey, and
D.~Levine, Phys.~Rev.~E {\bf 65}, 031304 (2002).

\bibitem{To95} S.~Torquato, Phys.~Rev.~Lett. {\bf 74}, 2156 (1995).

\bibitem{RT96} M.~D.~Rintoul and S.~Torquato,
J.~Chem.~Phys. {\bf 105}, 9258 (1996). 

\bibitem{Sp98} R.~J.~Speedy, Mol.~Phys. {\bf 95}, 169 (1998).

\bibitem{TTD00}
S.~Torquato, T.~M.~Truskett and P.~G.~Debenedetti,
Phys.~Rev.~Lett. {\bf 84}, 2064 (2000).

\bibitem{Torquato} S.~Torquato, {\it Random Heterogeneous Materials:
    Microstructure and Macroscopic Properties} (Springer-Verlag, New York, 2002).

\bibitem{DTS05} A.~Donev, S.~Torquato and F.~H.~Stillinger, Phys.~Rev.~E 
{\bf 71}, 011105 (2005).

\bibitem{XBH05} N.~Xu, J.~Blawzdziewicz and C.~S.~O'Hern,
Phys.~Rev.~E {\bf 71}, 061306 (2005).

\bibitem{HLLN02}
C.~S.~O'Hern, S.~A.~Langer, A.~J.Liu and S.~R.~Nagel,
Phys.~Rev.~Lett. {\bf 88}, 075507 (2002).

\bibitem{Luca05} L.~Angelani, G.~Foffi and F.~Sciortino,
cond-mat/0506447.

\bibitem{ConwaySloane} J.~H.~Conway and N.~J.~A.~Sloane, {\it Sphere Packings,
    Lattices and Groups} (Spriger-Verlag, New York, 1993).

\bibitem{Rogers} C.~A.~Rogers, {\it Packing and Covering}
(Cambridge University Press, Cambridge, 1964).


\bibitem{MP99} M.~M\'ezard and G.~Parisi,
J.~Chem.~Phys. {\bf 111}, 1076 (1999).

\bibitem{MP99b} M.~M\'ezard and G.~Parisi, Phys.~Rev.~Lett. {\bf 82},
747 (1999).

\bibitem{MP00} M.~M\'ezard and G.~Parisi, J.~Phys.:~Condens.~Matter {\bf 12},
6655 (2000).

\bibitem{CMPV99} B.~Coluzzi, M.~M\'ezard, G.~Parisi and P.~Verrocchio, 
J.~Chem.~Phys. {\bf 111}, 9039 (1999).

\bibitem{CFP98} M.~Cardenas, S.~Franz and G.~Parisi, J.~Phys.~A {\bf 31},
L163 (1998); J.~Chem.~Phys. {\bf 110}, 1726 (1999).

\bibitem{PZ05} G.~Parisi and F.~Zamponi, 
J.~Chem.~Phys. {\bf 123}, 144501 (2005).

\bibitem{PZ06a} G.~Parisi and F.~Zamponi,
J.Stat.Mech. (2006) P03017.

\bibitem{PTCC03}
M.~Pica Ciamarra, M.~Tarzia, A.~de~Candia and A.~Coniglio,
Phys.~Rev.~E {\bf 67}, 057105 (2003).

\bibitem{PZ06b} G.~Parisi and F.~Zamponi, cond-mat/0602661,
to appear on J.Stat.Phys. (2006)


\bibitem{MPV87} M.~M\'ezard, G.~Parisi and M.~A.~Virasoro, 
{\it Spin glass theory and beyond}
(World Scientific, Singapore, 1987).

\bibitem{KTW87b} T.~R.~Kirkpatrick and P.~G.~Wolynes, Phys.~Rev.~A {\bf 35}, 3072 (1987);
T.~R.~Kirkpatrick and D.~Thirumalai, Phys.~Rev.~Lett. {\bf 58}, 2091 (1987).

\bibitem{GM84} D.~J.~Gross and M.~M\'ezard, Nucl.~Phys.~B {\bf 240}, 431 (1984).

\bibitem{Mo95} R.~Monasson, Phys.~Rev.~Lett. {\bf 75}, 2847 (1995).

\bibitem{Hansen} J.-P.~Hansen and I.R.~McDonald, 
{\it Theory of simple liquids}
(Academic Press, London, 1986).

\bibitem{CAL05}
A.~Cavagna, A.~Attanasi and J.~Lorenzana,
Phys.~Rev.~Lett. {\bf 95}, 115702 (2005).

\bibitem{Kuunp} J.~Kurchan, private communication

\bibitem{He75} D.~Henderson, Mol.~Phys. {\bf 30}, 971 (1975).

\bibitem{Lu01} S.~Luding, Phys.~Rev.~E {\bf 63}, 042201 (2001).

\bibitem{SK00}
L.~Santen and W.~Krauth, Nature {\bf 405}, 550 (2000).

\bibitem{BW06} C.~Brito and M.~Wyart, cond-mat/0512197



\bibitem{FP99} H.~L.~Frisch and J.~K.~Percus, 
Phys.~Rev.~E {\bf 60}, 2942 (1999).

\bibitem{PS00} G.~Parisi and F.~Slanina, 
Phys.~Rev.~E {\bf 62}, 6554 (2000).

\bibitem{Tordlarge}
S.~Torquato and F.~H.~Stillinger, Phys.~Rev. E {\bf 73}, 031106 (2006); 
Journal of Experimental Mathematics, in press.

\bibitem{Zac}
K. Dawson, G. Foffi, M. Fuchs, W. G{\"o}tze, F. Sciortino, M. Sperl, P.
Tartaglia, Th. Voigtmann, and E. Zaccarelli,
Phys.~Rev.~E {\bf 63}, 01141 (2001);
F.~Sciortino, P.~Tartaglia and E.~Zaccarelli,
Phys.~Rev.~Lett. {\bf 91}, 268301 (2003);
F.~Sciortino, Nature Materials {\bf 1}, 145 (2002).

\bibitem{Velenich}
A.~Velenich, A.~Parola and L.~Reatto, cond-mat/0605466







\end{thebibliography}
\end{document}